# Towards the Socio-Algorithmic Construction of Fairness:

# The Case of Automatic Price-Surging in Ride-Hailing

Mateusz Dolata* and Gerhard Schwabe

*Department of Informatics, University of Zurich, Zurich, Switzerland*

* Binzmühlestrasse 14, 8050 Zurich, Switzerland, dolata@ifi.uzh.ch

https://orcid.org/0000-0002-2732-4465, https://www.linkedin.com/in/mateusz-dolata/

Biographies

Mateusz Dolata, born in 1988, is a full-time postdoctoral researcher at the University of Zurich's Department of Informatics. His research interests include the application of computer-supported collaboration and artificial intelligence for the greater good. He has co-authored several conference and journal articles that explore technological advancements' sociotechnical characteristics. He focuses on how people make sense of the latest technological phenomena.

Gerhard Schwabe, born in 1962, is a full professor at the University of Zurich's Department of Informatics. He studies information management and collaborative technologies. In multiple publications, he has focused on collaboration in workshops, city councils, advice-giving, and digital platforms. He addresses the impacts, application potentials, and management of novel technologies, such as tangible interfaces, blockchain, artificial intelligence, and human-robot collaboration.

# Towards the Socio-Algorithmic Construction of Fairness:
# The Case of Automatic Price-Surging in Ride-Hailing

Algorithms take decisions that affect humans, and have been shown to perpetuate biases and discrimination. Decisions by algorithms are subject to different interpretations. Algorithms' behaviors are basis for the construal of moral assessment and standards. Yet we lack an understanding of how algorithms impact on social construction processes, and vice versa. Without such understanding, social construction processes may be disrupted and, eventually, may impede moral progress in society. We analyze the public discourse that emerged after a significant (five-fold) price-surge following the Brooklyn Subway Shooting on April 12, 2022, in New York City. There was much controversy around the two ride-hailing firms' algorithms' decisions. The discussions evolved around various notions of fairness and the algorithms' decisions' justifiability. Our results indicate that algorithms, even if not explicitly addressed in the discourse, strongly impact on constructing fairness assessments and notions. They initiate the exchange, form people's expectations, evoke people's solidarity with specific groups, and are a vehicle for moral crusading. However, they are also subject to adjustments based on social forces. We claim that the process of constructing notions of fairness is no longer just social; it has become a socio-algorithmic process. We propose a theory of socio-algorithmic construction as a mechanism for establishing notions of fairness and other ethical constructs.

Subject classification codes: Artificial Intelligence Applications and Expert Systems; Empirical Studies of User Behavior; Information Systems – Decision Support Systems

# 1. Introduction

Many people experience injustice due to decisions taken by algorithms (Binns et al., 2018; Eubanks, 2017; Köchling et al., 2021; Marjanovic et al., 2021; O'Neil, 2016). However, a decision's impacts are not limited to specific subjects. We argue that decisions by algorithms at an individual or a group level can change what is considered fair or unfair in society. The decisions of organizations and governments have always been based on either implicit or explicit notions of fairness. These decisions later become reference points for future assessments of fairness. However, in contrast to human-made decisions, algorithms make decisions automatically, without considerations of legitimacy, transparency, competency, or impartiality, and achieve large-scale effects in a short time (Marjanovic et al., 2021; O'Neil, 2016). It may take years to discover, prove, and repair algorithmic flaws. However, before repairing is done, decisions taken by algorithms may form new or alter existing notions of fairness, i.e., the ethical construct of fairness, which serves as a reference point for an assessment of events and entities, and is constructed based on previous experiences and various cognitive and social-information processes. It is necessary to understand algorithms' roles in constructing notions of fairness so as to mitigate undesired shifts and social divides. Disadvantages that occur due to unfair algorithms must not become acceptable over time.

To understand algorithms' roles in constructing notions of fairness, we analyze social media conversations that emerged around the price-surge in ride-hailing after the Brooklyn Subway Shooting on April 12, 2022, in which 29 people were injured. Shortly after the shooting, the prices of ride-hailing offered by Uber and Lyft in New York City skyrocketed, becoming a hot topic in social media and on the news. The public discourse developed around the questions whether such surges are appropriate, justified, and fair, and who is better or worse off because of these price adjustments. Eventually, after several hours, the two ride-hailing companies disabled the price-surging routine in their algorithms and offered discounts or reimbursements. We claim that this discourse provides insights (though not a complete picture) into how people form notions of fairness through social exchanges and assess specific entities or decisions regarding fairness. Because the price-surging decision was automatic, it affected many people, and the companies reversed the decision post hoc, making this case particularly interesting to study social processes that emerge around decisions automatically made by algorithms.

 The collected data show that coming to an assessment is closely related to social exchanges, in which individuals not only present their opinions but also position themselves as members of a social group or are determined to be considered members of a group by others. This leads to the emergence of notions of fairness through solidarity with some participants or targets of the decision and antagonism toward other groups. Further, the fairness judgment also relates to the characteristics of the technology used in decision-making. In the ride-hailing case, the algorithm is a price-matching one. Participants who discuss price-surging practices often establish or present their positions by referring to algorithms' (assumed or real) limitations and abilities; these aspects are subject to collective interpretation during social exchanges. Accordingly, the algorithm becomes relevant in shaping a group's or an individual's notion of fairness. We seek to describe various roles algorithms take when people construct their interpretation of a situation.

This case shows that it is not only humans who help establish what is fair or unfair. Even though not active in social exchanges, algorithms become relevant shapers, for instance through voices that excuse or justify algorithms or that provide neutral or critical interpretations of their decisions. Algorithms set the stage for the discourse in the first place. To capture algorithms' impacts on framing the notions of fairness, we propose the concept of the socio-algorithmic construction of fairness, which extends the idea of the social construction of fairness (Alexander, 2002; Degoey, 2000) and follows the lines of social constructionism (Burr, 2015).

## 2. Fairness under Construction

### 2.1. *Algorithmic fairness*

The discourse on what is (un)fair goes back millennia (Miller, 2021). However, it has gained new momentum as probabilistic algorithms are now being used for taking high-stake decisions. In contrast to the original promise to provide impartial decisions, algorithms have turned out to be biased against underprivileged groups (Eubanks, 2017; Feuerriegel et al., 2020; O'Neil, 2016). To counterbalance this effect, the computer science community has proposed technical measures, including evaluation based on fairness metrics or counterfactual approaches (Dolata et al., 2022; Kordzadeh & Ghasemaghaei, 2021). While these solutions have improved the fairness of the outcomes for many prediction tasks, and the community keeps proposing new methods for specific application contexts and problems (Kordzadeh & Ghasemaghaei, 2021), it remains controversial whether algorithms will ever overcome biases.

However, the algorithmic fairness research is subject to many counterproductive assumptions (Dolata et al., 2022). For instance, improved algorithms are tested on generic, publicly available data sets, assuming the decisions' independence from the specific application context. A similar assumption affects the assessment of fairness by persons, for instance, to estimate an algorithm's fairness. Many researchers and practitioners assume that evaluating whether a decision (e.g., the distribution of rewards) is (un)fair is independent of the social and technical context in which this assessment is being made. Fairness is often tested through surveys in which subjects are asked to choose between distributions following various metrics of fairness (Grgic-Hlaca et al., 2018; Srivastava et al., 2019). This procedure abstracts from processes that lead to an assessment of fairness in 'normal' conditions. Further, much research abstracts from various ways in which fairness can be (re-)established and from procedural and informational aspects of fairness (Robert et al., 2020). We claim that whether a decision will ultimately be considered fair depends on social processes and on a decision's reception in a specific context (Colquitt, 2004; Mossholder et al., 1998; Tyler, 2021). Only if we understand how algorithmic decisions are treated in a social context can we address the issue of algorithmic fairness as a sociotechnical phenomenon (Dolata et al., 2022; Dolata & Schwabe, 2021).

## 2.2. *Social construction of fairness*

In the past, moral decisions and assessments were assumed to follow a logic expressed in the form of ethical theories (Bennett-Woods, 2005). Later, inspired by phenomenology and individualist trends in culture, notions of *personal construct psychology* and *radical constructivism* emerged; these focused on individual, internal, and private construction of ethical meanings. Finally, the concept of *social constructionism* emerged, and was applied to ethics and morality (Burr, 2015, 2018; Raskin & Debany, 2018). This notion relies on the assumption that "human constructions are not personal and private but, instead, are shared meanings we collaboratively create through our ongoing relationships with others" (Raskin & Debany, 2018, p. 345). Accordingly, "what is true, right, or good depends on the discourse from which one is operating" (Raskin & Debany, 2018, p. 345). Importantly, proponents of social constructionism argue against the criticism that misinterprets constructionism as an *anything goes* relativism. The emergence of meanings is limited by structures provided by human knowledge and experiences of the external world, as well as the structure provided by the social context (Burr, 2015; Raskin & Debany, 2018). At its core is the notion that knowledge (including ethical knowledge, such as assessments, norms, and moral concepts) is actively built and that humans build such knowledge to survive (Burr, 2018; Raskin & Debany, 2018). We claim that this perspective provides an adequate angle to understand how people establish meanings of fairness.

In parallel, organizational studies of fairness and justice proposed that the related notions come to exist through a social exchange within companies (Degoey, 2000; Lamertz, 2002; Lind et al., 1998). Such assessments emerge in a social and organizational context, and are sometimes triggered by these contexts (Degoey, 2000). Decisions are not taken in a vacuum; thus, fairness assessments should not be done in a vacuum.

Both lines of research propose various social mechanisms through which ethical notions are socially constructed. Degoey (2000) introduced the notions of *entity* and *event*. An entity can be understood as a decision-maker. By observing an entity's behaviors across various events, one establishes a model of how fair this entity is (Degoey, 2000). This model generates expectations on how this entity will behave in a specific event, i.e., how fair its decision will be. If a decision does not align with the model, one experiences a surprise and may need to adjust the model (Jones & Skarlicki, 2013; Skarlicki et al., 2015). The sources of information about an entity are not limited to the events one experiences; they also include reports from others, their interpretations, observations of how the entity deals with others, group discussions about events and entities, and emotions that others display about specific events or an entity (De Cremer et al., 2010; De Cremer & Van Hiel, 2010; Degoey, 2000; Geenen et al., 2013; Hollensbe et al., 2008; Jones & Skarlicki, 2013; Lind et al., 1998). All this information makes one engage in fairness assessment in the first place, but also induces solidarity or empathy (Hollensbe et al., 2008; Masterson & Tong, 2015). The processes will eventually lead to the emergence of a new notion of fairness as well as assessments of specific events or entities shared across a team, an organizational unit, or even society (Burr, 2018; Raskin & Debany, 2018).

Other mechanisms that impact on one's moral notions or assessments include specific conscious moves or communicative strategies. Rasking and Debany (2018) identified four mechanisms for general ethical construing: *moralizing* (imposing ethical constructs on others), *doubt* (having awareness of an upcoming change to one's ethical constructs), *confusion* (experiencing a dilemma that lies outside one's ethical construct system), and *crusading* (putting one's ethical constructs into practice to test them). Masterson and Tong (2015) identified specific moves that managers may use to impact on an employee's perceptions of justice and self-present as a fair entity. These moves include: justifications (refusing a decision's impropriety based on an objective), excuses (denial of responsibility), and various impression management types. Further mechanisms may include comparisons, classifications, building social relationships, or accounts of social identification (Lamertz, 2002). Finally, references to ethical theories and principles, such as utilitarianism or the ethics of care, can establish reference points and can shift one's expectations (Bennett-Woods, 2005).

Although the research into the social construction of ethical norms and assessments provides deep insights into construing activities among humans, they do not account for potential influences from algorithms or technology overall. However, technology may cause and moderate human responses to moral dilemmas (Verbeek, 2011). This moves them from passive recipients of human ethical assessments to active co-shapers of morally relevant behaviors and norms. Yet we lack an understanding of how algorithms' impacts intermingle with the social construction of fairness. We claim that algorithms are an element of the configuration in which fairness assessments are made, influencing them in fundamental ways. We explore various roles that algorithms can take in producing a fairness judgment about an event and related entities in a social context.

## 3. Price-Surging in Ride-Hailing

To analyze algorithms' impacts on constructing notions of fairness in social discourses, we investigated the exchanges about price-surges in ride-hailing services after the Brooklyn Subway Shooting on Tuesday, April 12, 2022. This occurred in the morning rush hour on a Manhattan-bound train in New York City. A masked passenger threw two smoke grenades and fired a handgun 33 times, injuring 29 passengers. The suspect was arrested one day later in Manhattan. The incident caused significant emergency responses to and disruption in various local transportation services, particularly in southern New York City. While the responsible transportation authority waived fees and increased frequency on selected lines to deal with the crisis, many commuters decided to stay at home or to use popular ride-hailing services for their travels instead. The prices of the ride-hailing services skyrocketed (increased many times compared to the same time on other days), not only in the direct surroundings of the incident, but also across the city. Many reports of the price increases were published on social media, launching a debate about price-surges and dynamic pricing practices.

We consider this case particularly interesting, for several reasons. First, it involves an algorithm that makes decisions automatically without the involvement of human actors.

Second, it was widely discussed in the news and on social media, providing sufficient material for analysis. A frequent aspect of the discourse was the fairness of the decision taken by the algorithms, with participants providing various arguments for or against the price-surge. Third, the companies, Uber and Lyft, overturned their algorithms' initial decision and reimbursed users who had paid higher prices. They announced their decisions on social media channels as responses to the public comments, which indicates that their reactions may have been caused by the public reactions rather than any company-internal insights. We will now describe the method we used to analyze the collected data and the results.

## *3.1. Method*

We used data from various sources so as to provide a multifaceted description of the discourse. Collected data included: (1) articles from LexisNexis and Factiva databases, (2) relevant threads from Reddit, and (3) relevant posts on Twitter. We applied the following query (adapted to each database): *brooklyn AND subway AND shooting AND (uber OR lyft)*. No additional filters were employed. We searched for news articles on May 7, 2022, and for Reddit and Twitter posts on May 10, 2022. For Twitter and Reddit, we manually selected additional posts and comments responding to or mentioning the ones found through the query. From the returned results, the following exclusion criteria were applied: (1) the item mentioned that the suspect had used Lyft to get to Brooklyn, (2) it only included a link to another item that was already included in the data set (e.g., retweets with no additional content), (3) it included independent references to Uber or Lyft and to the shooting, (4) it used the word 'uber' in the sense of *outstanding* or *supreme* (e.g., uber-conservative). The search, selection, and removal of duplicates resulted in 56 news and magazine articles and nine Reddit discussion threads, including 1,205 individual comments and 838 tweets (including commented-on retweets or responses to other tweets). The relevant news appeared between April 12 and 28; 43 tweets/posts (77%) on April 12 or 13. The relevant tweets/posts were published between April 12 and 26. Only 46 of all posts/tweets were published on April 13 or later, i.e., almost 98% were published within the first 40 hours after the shooting. The length of individual items varied strongly from tweets, including a couple of words, up to articles and longer comments of more than 2,500 words (though some of the longer news articles included an exact description of the shooting itself, apart from a discussion of the price-surge). Overall, the data set included spontaneous short comments as well as well-formulated and explained arguments.

A single coder applied a bottom-up coding procedure to the collected data, focusing on two core aspects. The coding schema involved 43 distinct categories; 31 were used to identify core events surrounding the price-surge along the timeline, document examples of price increases on the date of the shooting as well as on other occasions, and authorship of specific comments (e.g., if a post's author disclosed that they're an Uber driver). This yielded 142 segments with 217 codes assigned to them, forming a basis for the subsequent coding and analysis. On the other hand, the coder identified various opinions about the dynamic price increase on that day. He classified them according to (1) whether they were supportive or critical of the price increases and (2) arguments used to support the opinions. This resulted in 12 categories of arguments used by the discourse participants across 418 segments with 479 codes assigned to them. Overall, 547 individual segments were assigned 696 codes; 71 were coded as belonging to two or more categories. We then used the coded segments to identify various relevant groups affected by the algorithms' decisions, ethical concepts, and dilemmas expressed in the items, and the roles ascribed to the algorithms. This coding round adapted the concepts identified in the background literature on the social construction of fairness and on social constructionism.

The sensemaking of the collected and coded data evolved around an iterative process involving the authors and other interested researchers and non-researchers (see Acknowledgements). The sessions were used to assess the completeness and validity of the authors' interpretation of the collected data. The comments collected in this process illuminated the potential meanings of the identified and coded content. This procedure ensured that the presented results were not solely a product of an individual, subjective interpretation by the authors, but emerged in an intersubjective process involving actors with various backgrounds.

### *3.2. Results*

#### *3.2.1. Timeline*

Officially, the shooting occurred on April 12, 2022, at approximately 08:24 EDT. The three most widely commented posts reporting on the price-surge appeared online at 10:22[1], 10:43[2], and 11:04[3] and immediately caused many replies, and were reported in major online press services and were cited on Reddit. The first reaction by the raid-hailing companies was from Uber's Communications Manager, Freddi Goldstein, posted on her private Twitter account at 12:18.[4] Uber's official account or other channels provided no additional statement or declaration concerning the price-surge except a retweet of Goldstein's post at 17:26 entitled 'Statement from Uber.' The Uber

---

[1] https://twitter.com/shananigans022/status/1513885301793505284

[2] https://twitter.com/harharbinks/status/1513890452231143430

[3] https://twitter.com/harharbinks/status/1513890452231143430

[4] https://twitter.com/FreddiGoldstein/status/1513914512344330242

Support account used the same statement to respond to specific online comments. Goldstein's post was as follows:

> Our hearts go out to the victims of this morning's terrible shooting in Sunset Park. Uber has capped pricing citywide, and if anyone experienced unintended charges in the area during the emergency, we will work to refund them.

A statement from Lyft appeared on the company's official Twitter account as an indirect response to other posts at 13:54.[5] It reads:

> Thank you for making us aware of the situation on the ground. We've currently suspended prime-time pricing for riders in the area & are working to adjust fares for certain riders who paid prime-time prices when the situation first unfolded.

Further, Lyft has provided a longer blog post informing about the activation of a disaster response program that provided discounted rates for 3,000 users who used an adequate discount code.[6] The statements by Uber and Lyft did not directly address any of the complaints or comments on social media, nor did they engage in the assessment of the incident, except for Goldstein's calling the charges "unintended." There was no information on how many users were reimbursed by Uber and Lyft, and in what form(s). Further, the price reduction measures were not transparent, such that Twitter and Reddit users reported significantly higher ride prices across New York City, despite the announced measures.[7]

---

[5] https://twitter.com/lyft/status/1513938528245207041

[6] https://www.lyft.com/blog/posts/help-after-the-brooklyn-subway-shooting

[7] For instance, https://twitter.com/VintageHAJ/status/1514570279384752130 or
    https://twitter.com/Aliafonzy43/status/1514002385185746945

*3.2.2. Arguments*

Out of overall considered 2,043 comments, posts, and tweets from Twitter and Reddit, many did not directly discuss the price-surge decision. Still, they expressed emotional support for or disapproval of another post or of the general situation. The content was often brief, even limited to emoticons, such as "*Disgusting*" Also, many posts were retweets with additional content that simply summarized the original post's content, such as "*Uber saying they'll refund extra charges*" attached to a retweet of the original post announcing this action by Goldstein. Many newspaper articles simply reported the fact and cited relevant posts that we had already included in our data set. This explains why, despite the data set's size, only 547 individual, unique posts or article sections were identified as relevant for further analysis and were assigned 696 codes. In this subsection, we refer to results obtained from the analysis of 418 segments, assigned 479 codes, in which users argued why the price-surge was (un)justified in the context of the Brooklyn Subway Shooting.

The 418 statements we considered expressed the authors' opinions about whether the price-surge was justified, complemented by the reasoning behind their opinion. Overall, 218 statements saw the price-surge as justified and 193 as unjustified, while seven provided a mixed opinion. The mixed-opinion statements differentiated between the general practice of price-surges and this specific case. It is generally assumed that an algorithm automatically decides without human control. Table 1 summarizes the arguments concerning the specific decision to increase prices and exemplary statements from the data; it also indicates how often each category was used.

--- Table 1 here ---

*3.2.3. Stakeholders and dilemmas*

The collected statements indicate various stakeholders affected by the utilized algorithmic decisions. First, 167 statements (40% of the 418 considered here) mentioned various categories or riders: people who were directly affected by the shooting and wanted to swiftly leave the area, commuters who were affected by the (temporary) collapse of public transportation after the shooting, and generic riders who may belong to different social classes. Second, 98 statements (23.5%) focus on drivers who rely on ride-hailing platforms to earn money. Third, there are the companies that operate the platforms and employ the algorithms that make (automatic) decisions about the pricing and assignment between riders and drivers. Of the coded statements, 307 (73.5%) explicitly referred to the companies as affected entities; the companies were referred to in general terms or as algorithm developers or deciders. The authors of the collected statements tended to advocate for the interests of one of these stakeholders when discussing the price-related decisions' adequacy.

The comments indicated that there are potentials for various fairness dilemmas. First, what is a fair share for the company in the price paid by a rider? Second, what is a fair price for the offered service? Third, to what extent is it fair to limit the service's accessibility based on a rider's financial status? Fourth, to what extent is it fair to gain profits from an emergency, and in what circumstances? Fifth, who should benefit from and who should shoulder potential reimbursements or reduced pricing? Sixth, what is fair pay for a service in a high-risk area? All these dilemmas emerged from decisions taken ad-hoc by the algorithm: How much should a rider pay for a specific ride, and how much should the driver be paid for fulfilling a specific request? Both parties are informed about such amounts before agreeing on the service. However, additional rules are in place, such as the limitations on how many offers can be rejected by active drivers. However, not a single opinion or newspaper article covered all the aspects or provided a solution to these dilemmas.

Instead, the discourse participants implicitly chose an ethical standpoint that gives preference to one of these dilemmas and one of the affected groups. For instance, proponents of the underprivileged group's rights to transportation argued along the lines of an ethics of care. Those who affirmed companies' right to maximize their income referred to the law as the source of ethical guidance or to capitalism as a source of practice. Finally, those who suggested that companies should abstain from revenue-seeking during emergencies used the notion of virtues or principles. While it is unsurprising that people referred to various classic schools of ethics in their argumentation, 62 (15%) statements discussed the feasibility of potential adaptations to the algorithms on how these principles could be implemented. For instance, through automatic identification of emergency situations, forwarding the full price paid by the customers to the drivers, or even removing the possibility for drivers to enter high-risk areas so as to help protect their lives.

*3.2.4. Algorithm*

However, the algorithms were not only subject to suggestions for adaptations. In 69 statements (16.5%), commenters explicitly positioned algorithms as an independent entity responsible for automatic decisions – different to the company offering them. In 58 comments (14%), they were seen as an inherent product of the company and a key element of its business model used to realize its price politics. Of the comments, 152 (36.5%) presented the algorithm as implementing the more general market law of demand-and-supply. However, the algorithms were also presented as measures to motivate or even subjugate drivers (86 statements, 20.5%) or as tools that reinforce social differences (21 statements, 5%). Overall, the algorithms were portrayed in many different ways.

This made the users engage in various argumentative moves (cf. Table 1). Some users justified the decisions (automatically) taken by the algorithms by providing reasons for why the decisions were fair. Others sought to excuse the companies or algorithms by denying their responsibility for the price-surge. The advocates of the demand-and-supply law tended to moralize and sought to impose their ethical construct as the only possible one. Yet others expressed doubts and confusion, calling for a change in ethical constructs at the societal level. Finally, the companies took a very passive position on the discourse; however, the algorithms may be considered a tool for crusading, i.e., for actively testing their ethical constructs by implementing them in practice. The fact that the companies reversed the decisions (automatically) made by their algorithms (and without taking steps to prevent further situations such as the one presented here) shows that they accepted their algorithms as unfair and as having produced ethically problematic decisions.

### 4. Discussion

The analysis showed various trends. A slight majority of the analyzed statements supported the price-surge, while 46% were critical of it.[8] This shows a need and potential for broad societal discourse about the rules implemented in algorithms. Also, the various perspectives on the responsibility for the price-surge decisions are striking, and indicates a site of societal conflict. We will now discuss particular aspects and controversies uncovered in the data set.

---

[8] These numbers refer to the 418 statements, which were assigned 479 codes concerning their line of argumentation. The analysis of the overall set of 2,043 posts and 56 news articles may have generated different indications.

We started our research with the question about algorithms' roles in constructing notions of fairness. The results indicated that there is a bidirectional influence between the social discourse and the algorithmic decisions. Algorithms impact on the social discourse about what is (un)fair in various ways. First, they are an entity that makes a specific decision – just like managers are an entity that decides in an organizational context (Degoey, 2000). It means that people form opinions about fairness as a *trait* or an attribute of an algorithm and assess specific decisions of the algorithm as (un)fair events. The data showed that individual decisions about the price of a ride posted by several individuals caused strong discussion about the fairness of the algorithms Uber and Lyft used in their ride-hailing. Similar processes of forming a fairness reputation were reported in other instances of the social construction of justice (Lamertz, 2002; Lind et al., 1998). Because an algorithm's (automatic) decision becomes a reference point for the assessment of fairness and its calculations form an anchor value, it directly impacts on the formed notion of fairness. However, how discourse participants would react if the algorithm calculated, say, a hundred-fold increase as opposed to a five-fold one (as reported in our case) remains an open question.

Second, algorithms are subject to social expectations, like other decision-making entities. These expectations are based on knowledge about how technology works, as claimed by many statements in the data, or come from earlier experiences. Some comments that justify the price-surge very explicitly state that it would be naïve or dumb to expect algorithms to consider major social events or human emotions because this would be too hard to implement. Expectations, whether based on previous experiences or imposed through past social discourses, strongly impact on how fairness assessments of specific events are done (Geenen et al., 2013; Hollensbe et al., 2008;

Jones & Skarlicki, 2013). Specifically, say one takes the assumption that algorithms cannot reliably implement a notion of fairness that overlaps with human intuitions or needs; in this case, one is more likely to accept unfair decisions from an algorithm. Many people have the tacit understanding that algorithms cannot be as smart as humans in ethical issues and, therefore, silently accept unfair low-stake decisions. However, this acceptance may, with time, scale up and become a general agreement that unfair decisions form part of the process. This calls for the exact and multidimensional consideration of fairness in algorithmic decision-making, which should involve identifying all possibly affected stakeholders and all possible outcomes, rather than concentrating on certain aspects of the decision under consideration. For the ride-hailing case, this goes beyond the consideration of whether a price calculated for ride A is similar to the price calculated for a similar ride B. One must discuss fairness toward drivers, various rider categories, and the ride-sharing firms.

Third, algorithms create new solidarity. The observations indicate that many commenters expressed solidarity with specific stakeholders. This is similar to an organizational context, where some people may have solidarity with the decision-maker, and others will have solidarity with the subject affected by the decision (De Cremer et al., 2010; De Cremer & Van Hiel, 2010). However, there seemed to be no solidarity with the algorithm. Although algorithms are considered entities because they are a decision-maker, and are subject to expectations, they lack the property of being a de facto stakeholder in a situation. They are neither better nor worse off due to their own decisions, which minimizes the likelihood that people will have solidarity with them. Although statements in the category "the decision is taken automatically by an algorithm" see the algorithm most explicitly as an independent entity, not a single comment in the analyzed data expressed empathy with the algorithms. However, because humans are driven among others by the need to express solidarity when assessing an event's fairness (Skarlicki et al., 2015; Umphress et al., 2003), if they agree with the algorithm's decision, they need to have an addressee for their empathy or solidarity. This may move some participants to express solidarity with targets that were previously generally considered untrustful (e.g., big corporates or unlawful states) if they agree with some of the algorithm's decisions. This is reflected in comments that support Uber or Lyft's right to maximize their profit.

Fourth, algorithms are a vehicle for companies to implement their own (implicit or explicit) notion of fairness and test it in a real context while imposing it on society. In other words, companies employ algorithms for crusading, i.e., implementing ethical constructs (Raskin & Debany, 2018). Notably, the comments from Uber and Lyft did not acknowledge that the decision by their algorithm was unfair or wrong. They did not question the practice of price-surging – not even in this context. Only "charges in the area of emergency" were assessed as "unintended." The platform providers did not attend to the aspect of fairness or justice, neither in this specific case nor in their general terms and explanations given on the websites. During the social construction of fairness in an organizational context, decision-makers often engage in justifications and excuses concerning fairness. In the ride-hailing case, the justifications and excuses emerged in the public discourse and were produced by social media and news authors. Still, the companies remained completely silent on this matter. They present themselves as uninvolved in the event ("Thank you for letting us know...") and are at best able to react, but unable to act proactively. It remains an open question whether the ride-hailing platform providers consider their algorithms' (un)fairness at all.

However, we could observe an influence of the social discourse on the algorithm, as the outrage on social media caused the companies to reportedly change their algorithms' configuration and to reportedly offer additional discounts. One can see it as an algorithm adaptation so as to become aligned with the notion of fairness expressed on social media. While in this case the change was manual, through human intervention, and was limited in time and space, one could imagine algorithms that detect dissatisfaction and that try to adapt dynamically to market conditions and the ethical notions that emerge in a discourse. The adaptation of an algorithm (whether manually or automatically) follows the logic of sociotechnical systems in which social and technical systems mutually influence each other (Dolata et al., 2022; Dolata & Schwabe, 2021). The process of co-construing a notion of fairness or other ethical constructs involving autonomous algorithms and humans may be one of the mechanisms that will provide grounds for the functioning of advanced sociotechnical systems in the future.

This brings us to a controversy in the analyzed discourse: Who are the decision-makers behind the price-surging decisions? Some comments point to the algorithms that produce instance decisions; others point to the companies and/or developers who implemented a specific mechanism based on the law of demand-and-supply and associated rules (e.g., punishing drivers for refusing rides, the income distribution between the platform and its drivers). Yet statements that sought to justify the price-surge as an implementation of demand-and-supply logic or as a mechanism to motivate drivers, which amounted to more than one-third of the comments that supported the price-surge, barely touched on the question who is responsible. They seem to abstract from the entity that implements the rules and to associate them with the general socioeconomic order. This begs the question who is to blame.

Following Verbeek (2011), we claim that the moral responsibility for the price-surge is distributed among different social and technological entities. On the one hand, algorithms and other technologies (smartphones, GPS, cellular access to the Internet) give companies possibilities to impose dynamic pricing at scale. On the other hand, the companies use the algorithms to generate income and, in this case, also implement the algorithms and the dynamic pricing rules followed by the algorithms. However, also the customers who made choices for or against certain technologies (apps or platforms) took implicit moral decisions, accepting (or ignoring) the rules implemented in the algorithms. Thus, in addition to calls for normative ways to regulate technologies and their uses, it is essential to educate consumers about the consequences of their decisions and the risks of relying on automated decision-making.

## 5.     Towards Socio-Algorithmic Construction of Fairness

Both the results and the discussion indicate that we are experiencing a major change in how assessments concerning the fairness of specific events and entities are being made. Specifically, algorithms play key roles in the social process of constructing notions of what is (un)fair. They autonomously (and automatically) take decisions that are subject to fairness assessments, they are subject to fairness expectations and form expectations, they have the potential to induce new relationships of solidarity, and they propagate a specific notion of fairness such that the de facto protagonist of this notion can remain in the shadows. Finally, they change the distribution of moral responsibility. Thus, we claim that the process of constructing notions of fairness is no longer just social; it has become a socio-algorithmic process. This does not interrogate or overthrow basic assumptions behind the models of the social construction of fairness (Degoey, 2000; Hollensbe et al., 2008; Lamertz, 2002; Lind et al., 1998; Masterson & Tong, 2015) and, more generally, social constructionism in the context of ethics (Burr, 2015, 2018; Raskin & Debany, 2018). Rather, it proposes extending those models by acknowledging algorithms' active and growing roles in these processes. While we acknowledge that this idea needs further development and theorizing to turn it into a set of falsifiable or explanatory statements, in our view, the current study provides a solid point to initiate this enterprise.

This standpoint has major implications. First, it requires a more holistic understanding of material and technological forces in forming fairness in organizational and societal contexts. Second, it calls for a broader and longitudinal perspective on algorithmic fairness: while the current study identifies algorithms' roles in constructing notions of fairness, the question whether algorithms' impacts will increase remains open. Third, it requires the developers and users of algorithms to analyze and prepare for predictable exception events – not only the 'normal' operation. This will include a multidimensional analysis of potential moral dilemmas and the affected stakeholders. Finally, it provides new research directions for information systems and human-computer interaction communities: *How can algorithms actively and explicitly participate in construing a notion of fairness? How can they adapt to the notions that emerge in public discourses? How can algorithms' long-term impacts on moral/ethical dilemmas be made clear to the users? How can potential negative consequences of decisions be made transparent during the socio-algorithmic construction of fairness?*

**Acknowledgments**

We thank Robert Beaver, Robert O. Briggs, Xunyu Chen, Christian Dolata, and Carlos Palomares for their support during the analysis and interpretation of the collected data. Their inputs, suggestions, and field knowledge were a valuable resource to the authors throughout the entire process.

**Disclosure Statement**

The authors report that there are no competing interests to declare.

**Data Availability Statement**

The data were derived from the following resources available in the public domain: Twitter (https://twitter.com), Reddit (https://www.reddit.com), LexisNexis (https://www.lexisnexis.com), and Factiva (https://www.dowjones.com/professional/factiva). The coded and analyzed data are available, upon reasonable request, from the corresponding author, Mateusz Dolata.

**Funding Details**

The study was partially supported by the Swiss National Science Foundation under Grant 197485.

# References

Alexander, J. C. (2002). On the Social Construction of Moral Universals: The 'Holocaust' from War Crime to Trauma Drama. *European Journal of Social Theory*, *5*(1), 5–85. https://doi.org/10.1177/1368431002005001001

Bennett-Woods, D. (2005). *Ethics at a glance*. Regis University.

Binns, R., Van Kleek, M., Veale, M., Lyngs, U., Zhao, J., & Shadbolt, N. (2018). "It's Reducing a Human Being to a Percentage": Perceptions of Justice in Algorithmic Decisions. *Proceedings of the 2018 CHI Conference on Human Factors in Computing Systems*, 1–14. https://doi.org/10.1145/3173574.3173951

Burr, V. (2015). *Social constructionism* (Third edition). Routledge, Taylor & Francis Group.

Burr, V. (2018). Constructivism and the Inescapability of Moral Choices: A Response to Raskin and Debany. *Journal of Constructivist Psychology*, *31*(4), 369–375. https://doi.org/10.1080/10720537.2017.1384339

Colquitt, J. A. (2004). Does the Justice of the One Interact With the Justice of the Many? Reactions to Procedural Justice in Teams. *Journal of Applied Psychology*, *89*(4), 633–646. https://doi.org/10.1037/0021-9010.89.4.633

De Cremer, D., van Dijke, M., & Mayer, D. M. (2010). Cooperating when "you" and "I" are treated fairly: The moderating role of leader prototypicality. *Journal of Applied Psychology*, *95*(6), 1121. https://doi.org/10.1037/a0020419

De Cremer, D., & Van Hiel, A. (2010). 'Becoming Angry When Another is Treated Fairly': On Understanding When Own and Other's Fair Treatment Influences Negative Reactions. *British Journal of Management*, *21*(2), 280–298. https://doi.org/10.1111/j.1467-8551.2009.00653.x

Degoey, P. (2000). Contagious Justice: Exploring The Social Construction of Justice in Organizations. *Research in Organizational Behavior*, *22*, 51–102. https://doi.org/10.1016/S0191-3085(00)22003-0

Dolata, M., Feuerriegel, S., & Schwabe, G. (2022). A sociotechnical view of algorithmic fairness. *Information Systems Journal*, *32*(4), 754–818. https://doi.org/10.1111/isj.12370

Dolata, M., & Schwabe, G. (2021). How Fair Is IS Research? In S. Aier, P. Rohner, & J. Schelp (Eds.), *Engineering the Transformation of the Enterprise: A Design Science Research Perspective* (pp. 37–49). Springer International Publishing. https://doi.org/10.1007/978-3-030-84655-8_3

Eubanks, V. (2017). *Automating inequality: How high-tech tools profile, police, and punish the poor* (First Edition). St. Martin's Press.

Feuerriegel, S., Dolata, M., & Schwabe, G. (2020). Fair AI. *Business & Information Systems Engineering*, *62*(4), 379–384. https://doi.org/10.1007/s12599-020-00650-3

Geenen, B., Proost, K., Schreurs, B., van Dam, K., & von Grumbkow, J. (2013). What friends tell you about justice: The influence of peer communication on applicant reactions. *Revista de Psicología Del Trabajo y de Las Organizaciones*, *29*(1), 37–44. https://doi.org/10.5093/tr2013a6

Grgic-Hlaca, N., Redmiles, E. M., Gummadi, K. P., & Weller, A. (2018). Human Perceptions of Fairness in Algorithmic Decision Making: A Case Study of Criminal Risk Prediction. *Proceedings of the 2018 World Wide Web Conference*, 903–912. https://doi.org/10.1145/3178876.3186138

Hollensbe, E. C., Khazanchi, S., & Masterson, S. S. (2008). How Do I Assess If My Supervisor and Organization are Fair? Identifying The Rules Underlying Entity-Based Justice Perceptions. *Academy of Management Journal*, *51*(6), 1099–1116. https://doi.org/10.5465/amj.2008.35732600

Jones, D. A., & Skarlicki, D. P. (2013). How perceptions of fairness can change: A dynamic model of organizational justice. *Organizational Psychology Review*, *3*(2), 138–160. https://doi.org/10.1177/2041386612461665

Köchling, A., Riazy, S., Wehner, M. C., & Simbeck, K. (2021). Highly Accurate, But Still Discriminatory. *Business & Information Systems Engineering*, *63*(1), 39–54. https://doi.org/10.1007/s12599-020-00673-w

Kordzadeh, N., & Ghasemaghaei, M. (2021). Algorithmic bias: Review, synthesis, and future research directions. *European Journal of Information Systems*, 1–22. https://doi.org/10.1080/0960085X.2021.1927212

Lamertz, K. (2002). The social construction of fairness: Social influence and sense making in organizations. *Journal of Organizational Behavior*, *23*(1), 19–37. https://doi.org/10.1002/job.128

Lind, E. A., Kray, L., & Thompson, L. (1998). The Social Construction of Injustice: Fairness Judgments in Response to Own and Others' Unfair Treatment by Authorities. *Organizational Behavior and Human Decision Processes*, *75*(1), 1–22. https://doi.org/10.1006/obhd.1998.2785

Marjanovic, O., Cecez-Kecmanovic, D., & Vidgen, R. (2021). Algorithmic pollution: Making the invisible visible. *Journal of Information Technology*, 026839622110103. https://doi.org/10.1177/02683962211010356

Masterson, S. S., & Tong, N. (2015). Justice Perception Formation in Social Settings. In R. S. Cropanzano & M. L. Ambrose (Eds.), *The Oxford Handbook of Justice in the Workplace*. Oxford University Press. https://doi.org/10.1093/oxfordhb/9780199981410.013.13

Miller, D. (2021). Justice. In E. N. Zalta (Ed.), *The Stanford Encyclopedia of Philosophy* (Fall 2021). Metaphysics Research Lab, Stanford University. https://plato.stanford.edu/archives/fall2021/entries/justice/

Mossholder, K. W., Bennett, N., & Martin, C. L. (1998). A multilevel analysis of procedural justice context. *Journal of Organizational Behavior*, *19*(2), 131–141. https://doi.org/10.1002/(SICI)1099-1379(199803)19:2<131::AID-JOB878>3.0.CO;2-P

O'Neil, C. (2016). *Weapons of math destruction: How big data increases inequality and threatens democracy* (First edition). Crown.

Raskin, J. D., & Debany, A. E. (2018). The Inescapability of Ethics and the Impossibility of "Anything Goes": A Constructivist Model of Ethical Meaning Making. *Journal of Constructivist Psychology*, *31*(4), 343–360. https://doi.org/10.1080/10720537.2017.1383954

Robert, L. P., Pierce, C., Marquis, L., Kim, S., & Alahmad, R. (2020). Designing fair AI for managing employees in organizations: A review, critique, and design agenda. *Human–Computer Interaction*, *35*(5–6), 545–575. https://doi.org/10.1080/07370024.2020.1735391

Skarlicki, D. P., O'Reilly, J., & Kulik, C. T. (2015). The third-party perspective of (in)justice. In *The Oxford handbook of justice in the workplace* (pp. 235–255). Oxford University Press. https://doi.org/10.1093/oxfordhb/9780199981410.001.0001

Srivastava, M., Heidari, H., & Krause, A. (2019). Mathematical Notions vs. Human Perception of Fairness: A Descriptive Approach to Fairness for Machine Learning. *Proceedings of the 25th ACM SIGKDD International Conference on*

*Knowledge Discovery & Data Mining*, 2459–2468.

https://doi.org/10.1145/3292500.3330664

Tyler, T. R. (2021). Why People Obey the Law. In *Why People Obey the Law*. Princeton University Press. https://doi.org/10.1515/9781400828609

Umphress, E. E., Labianca, G. (Joe), Brass, D. J., Kass, E. (Eli), & Scholten, L. (2003). The Role of Instrumental and Expressive Social Ties in Employees' Perceptions of Organizational Justice. *Organization Science*, *14*(6), 738–753. https://doi.org/10.1287/orsc.14.6.738.24865

Verbeek, P.-P. (2011). *Moralizing technology: Understanding and designing the morality of things*. The University of Chicago Press.

**Table**

Table 1. Positions and argumentations for and against price-surging after the Brooklyn Subway Shooting, with examples. The number of occurrences refers to how often the code was assigned. The subheadings indicate an aspect of the decision, situation, or the algorithm addressed in the arguments listed beneath it.

| Position, argument, explanation, and number of occurrences | Example statements (original spelling and orthography) |
|---|---|
| **Algorithms' autonomy** | |
| The price-surge is **justified because the decision is made automatically by algorithms**. These commenters see the algorithms as independent from the firms that use them and assume that the algorithms did what they were programmed to do, such that no one is to blame for these decisions. Further, they claim that the decisions were unintentional – more like a mistake by the algorithm rather than a conscious decision by a person. Number of occurrences: 69 | "its not gouging. its an algorithm, like I said the more people that want uber rides, the higher the price will go." "Honestly I would defend uber in this situation, this is not their fault this happened. Noone manually adds surge, its an algorithm," "I highly doubt there was an intentional decision to try to surge it. Generally people attribute unforeseen side effects of algorithms to malice when it's just automation." "Why do people get angry at algorithms? It's an algorithm. It's not like some human was like: hey, fuck these particular people right here in brooklyn right now, I want to make an extra $20 per ride." |
| The price-surge is **unjustified because the algorithms are subject to human control**. These commenters see algorithms as purposefully designed to proliferate the moral and economic assumptions that underlie the firms' business models. They see the responsibility as being with the companies and the developers of the algorithms. Number of occurrences: 42 | "If the algos do this then there is something wrong with the algos. If they can't remove this bug from the algos then maybe this business model based on algos is no good." "Obviously it's an algorithm based on demand, but it is also unethical and potentially illegal. Seems fair to put some sort of control over this - public safety emergency overrides the rush algorithm." "'Computer says so' is not an excuse for profiteering. If the algorithm is setting prices based on demand without taking into account if it is a disaster or not, then the algorithm is wrong." |
| **Demand-and-supply logic** | |
| The price-surge is **justified because it reflects the** | "But no one is 'raising prices'. The prices are automatically adjusted by an algorithm based on supply and demand. It is the fairest method, both |
| **demand-and-supply logic of the market**.<br><br>These commenters assume that the law of demand-and-supply is fair and see no reason why this should be changed due to special circumstances. Further, they argue that it will motivate more drivers to take the rides, which will eventually reduce the demand and therefore the prices.<br><br>Number of occurrences: 88 | morally and economically. Is you disagree, what other method do you think would be more "morally fair"?"<br><br>"It's an algorithm based on supply and demand. After the attack, the demand for rideshare skyrocketed as people worked around the major subway delays. The algorithm sees this, and raises prices. Importantly, and what most people fail to understand, is that what the raised prices do is recruit more drivers to start taking rides. A guy who wasn't willing to drive for a $20 ride into Manhattan might be willing to get out of bed for a $40 ride into Manhattan. Capisce?" |
| Price-surging is **unjustified because demand-and-supply logic is inadequate in times of crisis** and because alternatives are possible.<br><br>These commenters claim that times of crisis may put random people in a disadvantaged position without any fault on their part. They also claim that demand-and-supply logic fails in some situations, for instance when supply is not flexible enough to cover the demand at all, or when there are additional limits imposed by external forces.<br><br>Number of occurrences: 76 | "i understand it, see my other comments. when supply is inelastic and you prioritize demand by raising prices, you are 1/ purely motivated by profit, and 2/ erring towards price gouging. the reason people are outraged isn't because they don't understand supply/demand dynamics, it is because they understand there is a price point where ride share companies know any price hike will not add drivers, but it will increase profits. it's a debatable behavior in an emergency, and it can be deemed price gouging."<br><br>"Yes, but also no. Uber has put a cap on how many drivers can work in New York. So it's not that it's just straight supply vs demand. Uber is artificially constructing the supply. (...) Uber restricted drivers in NYC because of an agreement with the yellow cabs to not totally put them out." |
| Business models of Uber and Lyft | |
| Price-surging is **justified because ride-hailing firms are for-profit organizations** and their primary goal is to maximize profit rather than to consider fairness or any other aspect of their decisions.<br><br>These commenters argue that privately-owned businesses cannot and should not take responsibility for events that are independent from their operations.<br><br>Number of occurrences: 21 | "Lmao what (...) are you talking about. Automatic surge pricing maximizes profits, which is the exclusive purpose of the business."<br><br>"If Uber was a public service funded by tax dollars, I'd agree, but it's a privately owned business designed to make money. They didn't set this up to operate in a disaster zone knowing they are taking advantage of a bad situation, and none of the people that were ordering Ubers after the shooting needed to order an Uber to leave the area, they did so by choice." |
| Price-surging is **unjustified because the ride-sharing firms' market practices and business models are too aggressive** and put parts of society or the economy at risk for the sake of companies' making money.<br><br>These commenters argue that privately-owned businesses need to be regulated to ensure better or fairer social and business environments. They also point out that ride-hailing firms benefit from higher prices, despite the fact that they provided no additional service and experienced no higher risk (in contrast to the drivers, who provided a service in a high-risk area).<br><br>Number of occurrences: 41 | "While I think anyone with a brain can figure out that the pricing system is automated, it's a great moment to really scrutinize their pricing model and questionable business practices. They've single handedly caused a decline in taxi supply, they've raised their price 50% yoy, and are planning on integrating taxis into their app Uber, Lyft and all the other car services need to be heavily regulated. They're not to big to fail. If their business model doesn't work, then it is what it is, but the government cannot sit by while they generate revenue by ripping consumers off and adding suspicious fees."<br><br>"The second fact is that Uber's (the company, I'm not talking about the drivers) profit margin increases when surge pricing in enacted. Note I'm not saying their gross profits, their profits per ride increases. That's because the cost of delivering a ride has not increased. It doesn't cost Uber more money to provide the individual service during a surge, but their price has still gone up. So when they enact surge pricing, it's to increase their profits."<br><br>"The parties benefiting from this type of surge pricing profiteering ina a crisis aren't solely the person taking the risks. Uber gladly takes their bump in money too. What is Uber risking in this equation?" |
| The difficulty of implementing alternatives | |
| Price-surging is **justified because it would be too hard to create an adequate algorithm** that accommodates unexpected events such as shootings.<br><br>These commenters seem to (implicitly) agree that the price-surging may be considered to be unfair, but it is still justified because the algorithm could not consider an outstanding situation. They frequently also refer to the decision of Uber and Lyft to reimburse the affected clients as a legitimate way to restore justice.<br><br>Number of occurrences: 17 | "The lack of people's awareness to how programmed systems are built is astounding. If you want to be able to cap uber prices, you have to build the manual overrides into the system. In advance."<br><br>"People don't understand that things that are easy for a human to do. Like realize you should not jack up prices as a result of demand created by a tragedy. Is very very difficult to program into a computer."<br><br>"The algorithm raised prices and then Uber capped prices and plans refunds but people still want to be mad because they didn't predict the attack and price accordingly." |
| Price-surging is **unjustified because it is possible to implement adequate measures** to prevent the algorithm from increasing prices in emergency situations.<br><br>These commenters claim that implementing technical or sociotechnical processes that would prevent unusually high price-surging is possible within the current state of technology and that it is the responsibility of ride-hailing firms to implement such measures.<br><br>Number of occurrences: 17 | "It seems like this would actually be almost trivial. Just have a checkbox when surge pricing that says 'There has been a civil emergency' that user can check. If a sufficiently large number of people check it, then drop the surge pricing until it is investigated. If people abuse it, ban them from the app. If they wanted to solve this, they absolutely could."<br><br>"Uber ended up refunding everyone's extra surge pricing charges anyway. They should have just not charged the higher prices in the first place. It wouldn't be hard to program the algorithm to notify an employee of an unexpected surge, so the employee can approve or deny it based on what is going on in that city." |
| Motivation and reimbursement of drivers | |
| Price-surging is **justified to motivate drivers and to reimburse them** for driving to a potentially dangerous area.<br><br>The commenters who used this argument saw the price-surge as a way to increase supply (like those who used the demand-and-supply argument) by attracting new drivers, but also stressed that it is fair to the drivers to pay them more for their willingness to drive to a high-risk area. Some participants in the discourse stated that they are Uber or Lyft drivers: they self- present as independent contractors and admit that increased prices motivate them to accept some riskier requests.<br><br>Number of occurrences: 57 | "The alternative is that you don't pay people more to risk their lives. That sounds like some dark, authoritarian bullshit to me. How is that not way more ethically and morally reprehensible than paying essentially hazard pay?"<br><br>"It's a good idea in theory. But it's fair for an Uber driver to demand more pay to go out in inclement weather, high congestion, late hours, or – in this case – the danger of an active shooting."<br><br>"Do uber drivers not deserve higher compensation for driving into a dangerous area? Are their lives less valuable than their passengers? Instead of asking Uber to turn off surge pricing, ask them to let drivers keep 100% of fares at these times."<br><br>"Tell me how are you going to incentivize more Uber drivers to head towards an area with active shooters and massive traffic jam? Are Uber drivers (independent contractors) supposed to do it purely out of goodness of their heart?" |
| Price-surging is **unjustified because it fails to motivate and reimburse the drivers** and sets inappropriate incentives.<br><br>These commenters claim that multiple mechanisms limit how much drivers benefit from a price-surge. Some who stated | "The average driver however, can still only do 3 rides per hour. On the individual driver level, even during this time of crisis (most likely some type of danger from violence, terrorism or weather) the driver isn't likely to earn additonal funds during the same hour that Uber's "earnings" shot up by 500%. On the macro level Uber's volume goes up. On the micro level, no, individual driver |
| that they are Uber or Lyft drivers complained that they barely get any benefit from price-surging, owing to these practices. Others also questioned the ethics of encouraging drivers to enter potentially dangerous areas through monetary incentives, because this puts the drivers who need money at greater risk than those who have some material security and who can choose to not drive or to not risk their status as a driver.<br><br>Number of occurrences: 30 | earnings don't go up because drivers are limited on time and space."<br><br>"I also drove Uber years back (...), however I was never once fairly compensated for surge pricing. Typically as a driver you would get a notification of a surge zone, and you could start heading that way. However you could only deny 2 riders in an hour before the app would kick you off for the day, possibly even suspend your driver account for longer. So what typically happened would be, you could deny the first rider on your way to the surge zone, and if a second rider was connected to you by the app before you made it to the surge zone you had to confirm their ride or else you were done for the day. You'd confirm their ride, which then may take you the opposite way of the surge, and by that time the surge is over with and you missed out."<br><br>"Uber drivers are not employees working for a fixed pay. Even if they are, it's questionable if the company should compel employees to work in a danger zone where an active shooter is still at large."<br><br>"This is terrible! But the other side of my brain that also did Uber driving as a side gig for a while ... those Uber drivers could unknowingly be picking up this shooter and helping him get away. That rate hike couldn't pay me enough to risk that without concealed carry." |
| The riders' economic status | |
| Price-surging is **justified because access to services should depend on economic power**.<br><br>These commenters argue **that** societies are naturally stratified based on people's wealth and that not everything can be provided to everyone, given the scarcity of the available resources.<br><br>Number of occurrences: 4 | "That IS RIGHT. That is how society determines the value of a person in relation to services/goods provided, via their ability to pay. How else would you? There are limited resources. "<br><br>"Uber shouldn't refund people who saw the price of a service and agreed to pay the price for said service. Like bro just don't get in the car?" |
| Price-surging is **unjustified because it may block poorer persons' access to a necessary service**, given that | "Many of the people using the subway in New York are doing so out of need, not want, and have no other affordable option to do so. When the subway is shut down, many people's only transportation options are to walk, take the bus, |
| they depend on it and/or that they cannot afford alternatives.<br><br>These commenters argue that it particularly affects poorer people who live in areas with worse public transportation, or who strongly depend on income from their existing job. It also puts poorer people in a worse position when they seek to leave a crime scene and puts them at greater risk.<br><br>Number of occurrences: 17 | or to take a cab/uber. (...) But if people can't affordably choose not to take [Uber or Lyft], then they are a captive market, which has moral implications. If you might lose your job for being late, what would you pay to keep it? $50? $200? $1,000?"<br><br>"Yes, but say we're talking about a terrorist situation and people are trying to flee. If prices surge 10x then only the wealthy would be able to get to safety. That's not right. We have laws against price gouging during crises."<br><br>"I can't believe you're really saying it's not a morally reprehensible position to incentivize regular folks to risk their lives to provide services to only those who can afford to pay exorbitant surge prices to escape an area in crisis. I'm not sure how that's any different than profiteering." |